\documentclass[aps,prd,nofootinbib,groupedaddress,twocolumn,floatfix]{revtex4}
\usepackage{graphicx}
\usepackage{epsfig}
\usepackage{dcolumn}
\usepackage{amsmath}
\def\Journal#1#2#3#4{{#1} {\bf#2}, #3 (#4)}
\def\PLB{{\rm Phys. Lett.}  B}
\def\PRL{\rm Phys. Rev. Lett.}
\def\PRD{{\rm Phys. Rev.} D}

\def\la{\langle}
\def\ra{\rangle}

\def\be{\begin{equation}}
\def\ee{\end{equation}}
\def\bea{\begin{eqnarray}}
\def\eea{\end{eqnarray}}
\input{epsfig.sty}
\begin{document}
%\vspace{2.in}
\title{PQCD analysis of exclusive $e^+e^-\to D_{[s]}^+D_{[s]}^-$ cross 
section at $\sqrt{s}=10.6$ GeV}
\author{ Ho-Meoyng Choi$^{a}$ and Chueng-Ryong Ji$^{b}$\\
$^a$ Department of Physics Education, Kyungpook National University,
     Daegu, Korea 702-701\\
$^b$ Department of Physics, North Carolina State University,
Raleigh, NC 27695-8202}
\begin{abstract}
We analyze the exclusive pseudoscalar $D_{[s]}^+D_{[s]}^-$
pair production in $e^+e^-$ annihilations at $\sqrt{s}=10.6$ GeV 
using a non-factorized PQCD with the
light-front wave function that goes beyond the peaking
approximation. We compare our non-factorized analysis
with the usual factorized analysis based on the peaking approximation
in the calculation of the cross section for the heavy meson pair production.
We also discuss the higher helicity contribution to the cross section.  
Our analysis provides a constraint on the size of quark transverse momentum
inside the $D$ meson from the recent Belle data,
$\sigma_{\rm Exp.}(e^+e^-\to D^+D^-)<0.04$.
\end{abstract}

%\pacs{13.40.Gp, 12.38.Lg, 13.20.He}
%\date{April 25, 2005}

\maketitle

\section{Introduction}
Recently, many theoretical works have been devoted to explain the large
discrepancy between theoretical and experimental results for the charmonium
production in $e^+e^-$ annihilations. For instance, the
data~\cite{Aubert,Abe,Uglov} for charmonium production cross sections
in $e^+e^-$ annihilations at the $B$-factory energy $\sqrt{s}=10.6$ GeV
differ a lot from the
theoretical predictions for both exclusive~\cite{Grozin,Bra1,Liu,Liu2,BKL2006}
and inclusive~\cite{Cho} processes 
although higher order corrections may reduce the differences~\cite{Chao2006}. 
A couple of years ago,
the Belle Collabotation~\cite{Uglov} reported the first measurement of the
$e^+e^-\to D^{*+}D^{*-}$ and $e^+e^-\to D^{+}D^{*-}$ cross sections and
polarizations at $\sqrt{s}\simeq 10.6$ GeV. They also set an upper limit
on the cross section for $e^+e^-\to D^+D^-$.
Interestingly, while the theoretical predictions based on the heavy quark
effective theory~\cite{Grozin} and the constituent quark model~\cite{Liu2}
for $e^+e^-\to D^{*+}D^{*-}$ and $e^+e^-\to D^{+}D^{*-}$ cross sections
are similar to the measured data~\cite{Uglov}, the predictions for
$e^+e^-\to D^+D^-$ cross section are either quite smaller~\cite{Grozin} or
somewhat larger~\cite{Liu2} than the data~\cite{Uglov}. 
%A particularly
%interesting feature of the QCD predictions first found by
%Brodsky and Ji~\cite{BJ} is the existence of a zero in the form factor
%for pseudoscalar meson pair production at the specific timelike value
%of $q^2$.

The above exclusive/inclusive meson pair productions provide a unique 
opportunity to investigate asymptotic behaviors of various meson form factors 
in the framework of perturbative quantum chromodynamics(PQCD). 
The heavy meson pair production is of special interest since gluons 
carrying large momentum transfers can be rather easily accessible in the 
kinematic region above the threshold. Also, the wavefunctions of heavy
systems may be well constrained due to the heaviness of constituents.
Thus, it has been pointed out that exclusive pair production 
of heavy mesons can be reliably predicted within PQCD~\cite{BJ}.

If the factorization theorem in PQCD is applicable to exclusive processes, then
the invariant amplitude for exclusive processes
factorizes into the convolution of the
valence quark distribution amplitude(DA) $\phi(x,q^2)$ with the hard scattering
amplitude $T_H$~\cite{BL}. To implement the factorization theorem at high 
momentum transfer, the hadronic wave function plays an important role linking
between long distance nonperturbative QCD and short distance PQCD. A 
particularly convenient and intuitive framework in applying PQCD to exclusive
processes is based upon the light-front(LF) Fock-state decomposition of hadronic 
state. 
%which arises naturally in the ``light-front(LF) quantization". 
In the LF framework, the valence quark DA is computed from the valence 
LF wave function 
$\Psi_n(x_i,{\bf k}_{\perp i})$
of the hadron at equal LF time $\tau=t + z/c$ which 
is the probability amplitude to find $n$ 
constituents(quarks,antiquarks, and gluons) with LF momenta 
$k_i=(x_i,{\bf k}_{\perp i})$ in a hadron. Here, $x_i$ and ${\bf k}_{\perp i}$
are the LF momentum fraction and the transverse momenta of the $i$th 
constituent in the $n$-particle Fock-state, respectively.

To lowest order in perturbation theory of the meson form factor calculation
at large momentum transfers,
the hard scattering amplitude $T_H$ is dominated by one-gluon exchange diagrams. 
For the factorization theorem to be applicable in the heavy meson
pair production analysis, the only consistent form of the
quark DA would be the  $\delta$ function, 
i.e.  $\phi(x,q^2)\sim\delta(x-m_Q/M)$ where $m_Q$ and $M$ 
are the heavy quark mass and the meson mass, respectively~\cite{JP}. 
In this so called ``peaking approximation", the momentum 
fraction carried out by $i$th constituent is equal to the ratio of the
constituent mass to meson mass, $x_i= m_i/M$. 
This relation implies $M=m_1 + m_2$, i.e. zero-binding energy limit.

However, as pointed out in Ref.~\cite{JP}, if the quark DA is not an 
exact $\delta$ function, i.e. 
${\bf k}_\perp$ in the soft bound state LF wave function 
can play a significant role,
the factorization theorem is no longer applicable. To go
beyond the peaking approximation, the invariant amplitude
should be expressed in terms of the LF wave function 
$\Psi(x_i,{\bf k}_{\perp i})$ rather than the quark DA.
In Ref.~\cite{JP}, the validity issue of peaking approximation 
for the heavy meson pair production processes was discussed
using the LF model wave function $\Psi(x_i,{\bf k}_{\perp i})
\propto \exp(-M^2_0/\beta^2)$,
where $M_0$ is the invariant mass of the constituent quark and antiquark 
defined by $M^2_0=\sum_i({\bf k}^2_\perp + m^2_i)/x_i$ and
$\beta$ is the gaussian parameter.
The limit $\beta\to 0$ corresponds to 
the peaking approximation(i.e. zero-binding energy limit $M=m_1+m_2$). 
In the analysis of the heavy-heavy system like $B_c(b{\bar c})$ meson, it 
was 
found that the effect of
going beyond the peaking approximation ($\beta$ up to 100 MeV) was not 
important compared to the peaking
approximation limit(i.e. $\beta\to 0$)~\cite{JP}.
However, it is not yet clear if the same conclusion would apply to
the heavy-light system such as $D$ and $B$ mesons.
Moreover, the initial analysis limited only up to 
$\beta \leq 100$ MeV may not be sufficient to draw a definite
conclusion on the validity of the peaking approximation.

The main purpose of this work is to extend the previous analysis~\cite{JP}
and point out that the recent Belle data~\cite{Uglov} can provide a rather stringent 
constraint on how broad or narrow the $D^\pm$ meson quark DA is. Clarifying the relation 
between the $\beta$ value and transverse momentum, ${\bf k}_\perp$, is 
a particularly important issue since the quark DA is
very sensitive to the $\beta$ value and the different shape of the quark
DA could enhance or reduce the cross section for the exlcusive meson pair
production in $e^+ e^-$ annihilations. Incidentally, 
Bondar and Chernyak~\cite{BC} considered a rather broad
quark DA(i.e. rather significant binding energy effect) instead of 
$\delta$-type quark DA to explain the data for the exclusive
$e^+e^-\to J/\psi + \eta_c$ process.  Ma and Si~\cite{MS} also
previously discussed the variation of DA to explain the data
for the same process. 

In this work, we stress a consistency of our analysis in going beyond the
peaking approximation. In particular, we confirm that the $\beta$ value in  
our model LF wave function is related with the transverse momentum via 
$\beta_{Q\bar{Q}}=\sqrt{\la{\bf k}^2_\perp\ra}_{Q\bar{Q}}$.
As expected, the non-zero $\beta$ value corresponds to the transverse 
size of the meson and $\beta\to 0$ limit corresponds to the peaking
approximation(i.e. zero binding energy limit) as discussed in~\cite{JP}. 
This implies that it may be significant to keep the transverse momentum
${\bf k}_\perp$ both in the wavefunction part and the hard scattering part
together before doing any integration in the amplitude if $\beta$ is not
so close to zero or the binding energy effect is not negligible.
Thus, we think that the factorization of amplitude
by integrating out the transverse momentum seperately 
in the wavefunction part and in the hard scattering part 
may not provide a consistent analysis to take into account the 
binding energy effect. This could distinguish our method from Ref.\cite{BC}
to take into account the binding energy effect.

We also note that our gaussian parameter $\beta$ is not 
chosen arbitrarily but fixed by the variational principle for the well-known 
linear plus Coulomb interaction motivated by QCD~\cite{CJ2}, 
which in turn uniquely determine the shape of the quark DA 
in our model calculation. This implies that the recent data by the
Belle collaboration~\cite{Uglov} provide a useful test
on our model calculation.
%While the higher helicity contribution to
%form factor was neglected in~\cite{JP}, we consider this effect in this 
%work. 

The paper is organized as follows. 
In Sec. II, we describe the formulation of our 
light-front quark model (LFQM), which has been quite
successful in describing the static and non-static properties of the 
low- lying mesons~\cite{CJ2,CJ99}.  
In Sec. III, the transverse momentum dependent
hard scattering amplitude for the meson is given within the LF framework. 
The contribution to the
meson form factor from higher helicity components is also given in this 
section.  
In Sec.IV, the analytic continuation from spacelike region to the timelike
region is introduced to obtain the cross section for the pseudoscalar
meson pair($M\bar{M}$) production in $e^+e^-$ annihilations. As a validity
check of our model, we also show that our result for the meson form factor 
obtained in Sec.III reduces to the peaking approximation 
in the $\beta\to 0$(i.e. zero-binding) limit. 
In Sec. V, we present the numerical results for the
$e^+e^-\to D^+_{[s]}D^-_{[s]}$ cross section and compare 
with the available data. 
Summary and conclusions follow in Sec. VI.
In the Appendix, we briefly summarize our proof of vanishing
contribution from the light-front gauge part in $M=M_0$ limit.

%{\label{sect.I}}

\section{Model Description}
In our LFQM, 
the meson wave function is given by
\bea\label{wf}
\Psi_M^{JJ_z}(x,{\bf k}_\perp,\lambda\bar{\lambda})
&=& \phi(x,{\bf k}_\perp)
{\cal R}^{JJ_z}_{\lambda\bar{\lambda}}(x,{\bf k}_\perp).
\eea
where $\phi(x,{\bf k}_\perp)$ is the radial wave function and 
${\cal R}^{JJ_z}_{\lambda\bar{\lambda}}(x,{\bf k}_\perp)$ is the spin-orbit
wave function obtained by the interaction-independent Melosh 
transformation~\cite{Melosh}
from the ordinary equal-time static spin-orbit wave function assigned by
the quantum numbers $J^{PC}$. The meson wave function in Eq.~(\ref{wf}) is
represented by the Lorentz-invariant variables $x_i=p^+_i/P^+$, 
${\bf k}_{\perp i}={\bf p}_{\perp i}- x_i{\bf P}_\perp$ and $\lambda_i$, 
where $P,p_i$ and $\lambda_i$ are the meson momentum, the momentum and the
helicity of the constituent quarks, respectively.

The radial wave function $\phi(x,{\bf k}_\perp)$ of a ground state 
pseudoscalar meson($J^{PC}=0^{-+}$) is given by
\bea\label{phi}
\phi(x,{\bf k}_\perp)&=&\biggl(\frac{1}{\pi^{3/2}\beta^3}\biggr)^{1/2}
\exp(-{\vec k}^2/2\beta^2),
\eea
where the gaussian parameter $\beta$ is related with the size of the meson.
Here, the longitudinal component $k_z$ of the three momentum is given
by $k_z=(x_1-\frac{1}{2})M_0 + (m^2_2-m^2_1)/2M_0$ with the invariant mass
\bea\label{M0}
M^2_0&=& \frac{{\bf k}^2_\perp + m^2_1}{x_1}
+\frac{{\bf k}^2_\perp + m^2_2}{x_2},
\eea 
where $x_1=x$ and $x_2=1-x$.
The covariant form of the  spin-orbit wave function 
${\cal R}^{00}_{\lambda\bar{\lambda}}(x,{\bf k}_\perp)$ for 
the pseudoscalar meson is given by
\bea\label{spin_cov}
{\cal R}^{00}_{\lambda\bar{\lambda}}=
-\frac{\bar{u}(p_1,\lambda)\gamma_5 v(p_2,\bar{\lambda})}
{\sqrt{2}[M^2_0-(m_1-m_2)^2]^{1/2}},
\eea
and its explicit matrix form is given by
\begin{eqnarray}\label{so}
{\cal R}^{00}_{\lambda\bar{\lambda}}=\frac{1}{C}
\left(\begin{array}{cc}
-k^L & x_1m_2 + x_2m_1 \\
-x_1m_2-x_2m_1 & -k^R 
\end{array} \right),
\end{eqnarray}
where $C=\sqrt{2x_1x_2[M^2_0-(m_1-m_2)^2]}$ and 
$k^{R(L)}=k_x\pm ik_y$. Note that 
$\sum_{\lambda\bar{\lambda}}R^{00\dagger}_{\lambda\bar{\lambda}}
R^{00}_{\lambda\bar{\lambda}}=1$.
The normalization of our wave function is given by
\bea\label{norm}
\sum_{\lambda\bar{\lambda}}&&\hspace{-0.5cm}
\int d^3 k|\Psi^{00}_M(x,{\bf k}_\perp,\lambda\bar{\lambda})|^2
\nonumber\\
&=&\int^1_0 dx
\int d^2{\bf k}_\perp\biggl(\frac{\partial k_z}{\partial x}\biggr)
|\phi(x,{\bf k}_\perp)|^2 = 1,
\eea
where the Jacobian of the variable transformation $\{x,{\bf k}_\perp\}\to
\vec{k}=({\bf k}_\perp,k_z)$ is given by
\bea\label{Jacob}
\frac{\partial k_z}{\partial x}=
\frac{M_{0}}{4x_1x_2}\biggl\{
1-\biggl[\frac{(m_1-m_2)^2}{M^2_{0}}\biggr]^2\biggr\}.
\eea
The effect of the Jacobi factor has been analyzed in 
Ref.~\cite{CJ_Jacob}.

With this normalization, the root-mean-square(r.m.s.) value of the transverse 
momentum($\sqrt{\la{\bf k}^2_\perp\ra_{Q\bar{Q}}}$) is
obtained via 
\bea\label{rmsT}
\la{\bf k}^2_\perp\ra_{Q\bar{Q}} &=&
e_Q\int d^3k |{\bf k}^2_\perp||\phi(x,{\bf k}_\perp)|^2 
+ (Q\leftrightarrow\bar{Q}).
\eea
Numerically, we confirm that 
$\sqrt{\la{\bf k}^2_\perp\ra_{Q\bar{Q}}}=\beta_{Q\bar{Q}}$.
The numerical values of $\beta_{Q\bar{Q}}$ are discussed
in Sec. V (see Table I).

The quark distribution amplitude(DA) of a meson, $\phi_{M,\lambda}(x,Q)$, 
i.e. the probability of
finding collinear quarks up to the scale $Q$ in the $L_z=0$($s$-wave) 
projection of the meson wavefunction~\cite{BL} is defined by
\bea\label{Dis}
\phi_{M,\lambda}(x,Q)&=&\int^Q [d^2{\bf k}_\perp]
\Psi(x,{\bf k}_\perp,\lambda\bar{\lambda}),
\eea
where 
%$[d^2 {\bf k}_\perp]=d^2 {\bf k}_\perp/16\pi^3$ for 
%$\Psi=\Psi^{00}_{\rm BHL}$ and  
$[d^2 {\bf k}_\perp]=d^2 {\bf k}_\perp\sqrt{\partial k_z/\partial x}
/\sqrt{16\pi^3}$ for $\Psi=\Psi^{00}_M$.
%, respectively.

\section{Hard scattering amplitude with ${\bf k}_\perp$-dependence}
In this section, we calculate the pseudoscalar meson electromagnetic form 
factor in the region where the PQCD is applicable. 
Our calculation is carried out using the 
Drell-Yan-West frame~\cite{DYW}($q^+=q^0+q^3=0$)
with ${\bf q}^2_\perp=Q^2=-q^2$.  
The momentum assignment in the $q^+=0$ frame is given by
\bea\label{km1}
P&=&(P^+, \frac{M^2}{P^+},{\bf 0}_\perp),\;
P'= (P^+, \frac{M^2+{\bf q}^2_\perp}{P^+},{\bf q}_\perp)
\nonumber\\
q&=&(0,\frac{{\bf q}^2_\perp}{P^+},{\bf q}_\perp),
\eea
where prime denotes the final state momentum and $q=P'-P$ 
and $M$ is the physical meson mass.

As a starting point, the electromagnetic form factor
of a pseudoscalar meson is given by
a convolution of initial and final meson wavefunctions:
 
\bea\label{F_NPQCD}
F^{\rm soft}_M(Q^2)&=&\sum_{\lambda\bar{\lambda}}\sum_{j}e_j \int d^3k
\Psi^{00*}_M(x,{\bf k'}_\perp,\lambda\bar{\lambda})
\nonumber\\
&&\times \Psi^{00}_M(x,{\bf k}_\perp,\lambda\bar{\lambda}),
\eea
where $d^3k= dx d^2{\bf k}_\perp
\sqrt{\partial k_z/\partial x}$,
${\bf k'}_\perp= {\bf k}_\perp + x_2{\bf q}_\perp$ 
and $e_j$ is the electric charge of the struck quark. 
%Note that the
%corresponding measure for the BHL-type wave function is given by 
%$[d^3k]_{\rm BHL}= dx d^2{\bf k}_\perp/16\pi^3$ according 
%to Eq.~(\ref{A_const}). 

At high momentum transfers, the meson form factor   
can be calculated within the leading order PQCD 
by means of a homogeneous Bethe-Salpeter equation
for the meson wavefunction. 
Taking the perturbative kernel of the Bethe-Salpeter equation as a 
part of hard
scattering amplitude $T_H$,
one can get the meson electromagnetic
form factor given by\footnote{ We should note that the corresponding
measure $[d^3k d^3l/16\pi^3]$ in Eq.~(\ref{F_PQCD}) has to be replaced
by $[dxd^2{\bf k}_\perp/16\pi^3][dyd^2{\bf l}_\perp/16\pi^3]$
for the BHL-type wave wave function~\cite{BHL}.  }
\begin{eqnarray}\label{F_PQCD}
F^{\rm Hard}_M(Q^2)&=&
\int \frac{d^3k d^3l}{16\pi^3} \Psi^{00*}_M(y,{\bf l}_\perp)
T_H(x,y,{\bf q}_\perp,{\bf k}_\perp,{\bf l}_\perp)
\nonumber\\
&&\times \Psi^{00}_M(x,{\bf k}_\perp)
\nonumber\\
&=&\int \frac{d^3k d^3l}{16\pi^3}\phi(y,{\bf l}_\perp)
{\cal T}_H
\phi(x,{\bf k}_\perp),
\end{eqnarray}
where $T_H$ contains all two-particle irreducible amplitudes
for $\gamma^* + q{\bar q}\to q{\bar q}$ from the iteration
of the LFQM wavefunction with the Bethe-Salpeter kernel.
In the 2nd line of Eq.~(\ref{F_PQCD}), we combined the spin-orbit 
wave function into the original $T_H$ to form a new ${\cal T}_H$, i.e.
\bea\label{TH}
{\cal T}_H = {\cal R}_0 T_H^{(\lambda+\bar{\lambda}=0)}
+ {\cal R}_{\pm 1} T_H^{(\lambda+\bar{\lambda}=\pm1)},
\eea
where 
\bea\label{spin}
{\cal R}_0 &=& {\cal R}^*_{\uparrow\downarrow}(y,{\bf l}_\perp)
{\cal R}_{\uparrow\downarrow}(x,{\bf k}_\perp)
+ {\cal R}^*_{\downarrow\uparrow}(y,{\bf l}_\perp)
{\cal R}_{\downarrow\uparrow}(x,{\bf k}_\perp)
\nonumber\\
&=& 2\frac{[y_1 m_2 + y_2m_1][x_1 m_2 + x_2m_1]}
{C_{0x}C_{0y}},
\nonumber\\
{\cal R}_{\pm1} &=& {\cal R}^*_{\uparrow\uparrow}(y,{\bf l}_\perp)
{\cal R}_{\uparrow\uparrow}(x,{\bf k}_\perp)
+ {\cal R}^*_{\downarrow\downarrow}(y,{\bf l}_\perp)
{\cal R}_{\downarrow\downarrow}(x,{\bf k}_\perp)
\nonumber\\
&=& 2\frac{{\bf k}_\perp\cdot {\bf l}_\perp}
{C_{0x}C_{0y}},
\eea
with $C_{0x}=C$ and $C_{0y}=C(x\leftrightarrow y, 
{\bf k}\leftrightarrow{\bf l}_\perp)$ [see below Eq.~(\ref{so}) for $C$]. 
The hard scattering amplitudes
$T_H^{(\lambda+\bar{\lambda}=0)}$ and 
$T_H^{(\lambda+\bar{\lambda}=\pm 1)}$ in Eq.~(\ref{TH}) represent the
contributions from the ordinary-helicity and higher-helicity components,
respectively. 
%In conventional PQCD analysis, the hard scattering amplitude
%$T_H^{(\lambda+\bar{\lambda}=0)}$ for the pion form factor in leading twist
%has been obtained by 
%\bea\label{TH_pi}
%T_H^{(\lambda+\bar{\lambda}=0)}&=&\frac{16\pi \alpha_s C_F}{Q^2}
%\biggl(\frac{e_u}{x_2y_2}+\frac{e_{\bar{d}}}{x_1y_2}\biggr),
%\eea
%However, the transverse momentum ${\bf k}_\perp$-effect
%in the hadronic wave function has been known to lead to a substantial
%contribution to the PQCD contribution to the pion form factor~\cite{JK,HW}.
%We shall analyze PQCD contribution to the heavy meson 
%including the transverse momentum not only in the hard scattering kernel 
%but also in the LF wave function to extend the pion case. We also discuss
%the higher helicity contribution to the hard scattering amplitude for the
%heavy meson case.

\begin{figure}[t]
\includegraphics[width=3.2in,height=2.5in]{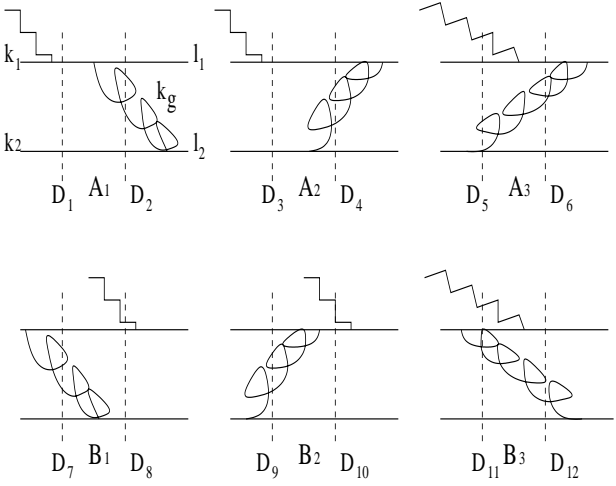}
\caption{Leading order light-front time-ordered diagrams for the meson
form form factor.}
\label{fig1}
\end{figure}

To lowest order in perturbation theory, the hard scattering amplitude 
$T_H(x,y,{\bf k}_\perp,{\bf l}_\perp)$ is  
calculated from the time-ordered one-gluon-exchange diagrams
shown in Fig.~\ref{fig1}.  The internal momenta for $(+,\perp)$-components 
are given by

\bea\label{km2}
k_1&=& (x_1 P^+_1, {\bf k}_\perp),\;
k_2= (x_2 P^+_1, -{\bf k}_\perp)\nonumber\\
l_1 &=& (y_1P^+_1, y_1{\bf q}_\perp + {\bf l}_\perp),\;
l_2=(y_2 P^+_1, y_2{\bf q}_\perp - {\bf l}_\perp).
\nonumber\\
\eea
In each diagram in Fig.~\ref{fig1}, the instantaneous diagrams for the 
intermediate constituents are included using the technique shown in 
Ref.~\cite{BL}. In the LF gauge $A^+=0$, the gluon propagator is given by
\bea\label{glue}
d_{\mu\nu}=-g_{\mu\nu} + \frac{(k_g)_\mu\eta_\nu + (k_g)_\nu\eta_\mu}{k^+_g},
\eea
where $\eta^+=0,\eta^-=1$, and ${\vec\eta}_\perp=0$.

Hard scattering amplitudes for the helicity 
$\lambda+\bar{\lambda}$(=0 or $\pm1$) 
components for the diagrams $A_i(i=1,2,3)$ are given by 
\bea\label{eq1}
T_{A_1}^{(\lambda+\bar{\lambda})}&=& \frac{\theta(y_2-x_2)}{y_2-x_2}
\frac{N_A^{(\lambda+\bar{\lambda})}+N_{A_1}}{D_1 D_2},\nonumber\\
T_{A_2}^{(\lambda+\bar{\lambda})}&=& \frac{\theta(x_2-y_2)}{x_2-y_2}
\frac{N_A^{(\lambda+\bar{\lambda})}+N_{A_2}}{D_3 D_4},\nonumber\\
T_{A_3}^{(\lambda+\bar{\lambda})}&=& \frac{\theta(x_2-y_2)}{x_2-y_2}
\frac{N_A^{(\lambda+\bar{\lambda})}+N_{A_3}}{D_5 D_6},
\eea
where the energy denominators are given by
\bea\label{eq4}
D_1&=& M^2 + {\bf q}^2_\perp
- \frac{({\bf k}_\perp + {\bf q}_\perp)^2+m^2_1}{x_1}
-\frac{{\bf k}^2_\perp + m^2_2}{x_2},
\nonumber\\
D_2&=& M^2+ {\bf q}^2_\perp
- \frac{(y_1{\bf q}_\perp + {\bf l}_\perp)^2+m^2_1}{y_1}
-\frac{{\bf k}^2_\perp + m^2_2}{x_2}
\nonumber\\
&&-\frac{(y_2{\bf q}_\perp + {\bf k}_\perp - {\bf l}_\perp)^2}{y_2-x_2},
\nonumber\\
D_3 &=& D_1,
\nonumber\\
D_4&=& M^2+ {\bf q}^2_\perp
- \frac{({\bf k}_\perp + {\bf q}_\perp)^2+m^2_1}{x_1}
\nonumber\\
&&-\frac{(y_2{\bf q}_\perp + {\bf k}_\perp - {\bf l}_\perp)^2}{x_2-y_2}
-\frac{(y_2{\bf q}_\perp - {\bf l}_\perp)^2 + m^2_2}{y_2},
\nonumber\\
D_5 &=& M^2 - \frac{{\bf k}_\perp^2 + m^2_1}{x_1}
- \frac{(y_2{\bf q}_\perp + {\bf k}_\perp - {\bf l}_\perp)^2}{x_2-y_2}
\nonumber\\
&&- \frac{(y_2{\bf q}_\perp - {\bf l}_\perp)^2 + m^2_2}{y_2},
\nonumber\\
D_6 &=& D_4.
\eea
The common $N_A$ in Eq.~(\ref{eq1}) is obtained from the Feynman 
gauge($g^{\mu\nu}$) part and given by
\bea\label{eq2}
N_{A}^{(0)}&=&\frac{8}{x_1x_2y_1y_2}
\biggl[ x^2_2y_1y_2 {\bf q}^2_\perp + x_1x_2 {\bf l}^2_\perp
+ y_1y_2{\bf k}^2_\perp 
\nonumber\\
&&+ x_2(x_1y_1+x_2y_2){\bf l}_\perp\cdot {\bf q}_\perp
+ 2x_2y_1y_2 {\bf k}_\perp\cdot {\bf q}_\perp 
\nonumber\\
&&+ (x_1y_1 + x_2y_2) {\bf k}_\perp\cdot {\bf l}_\perp
+ x_1y_1 m^2_2 + x_2y_2 m^2_1 
\nonumber\\
&&+ m_1m_2(y_1-x_1)(y_2-x_2) \biggr],
\nonumber\\
N_{A}^{(\pm 1)}&=&\frac{8
[ {\bf k}_\perp\cdot {\bf l}_\perp + x_2 {\bf l}_\perp\cdot {\bf q}_\perp
+ x_2y_2 m^2_1 + x_1y_1 m^2_2]}{x_1x_2y_1y_2},
\nonumber\\
\eea
where the last mass term $m_1m_2(y_1-x_1)(y_2-x_2)$ in
$N_A^{(\lambda+\bar{\lambda}=0)}$ comes from the helicity flip contribution. 
In Eq.~(\ref{eq1}),
$N_{A_i}(i=1,2,3)$ are obtained from the LF gauge parts proportional
to $1/k^+_g$ and given by
\bea\label{eq3}
N_{A_1}&=& \frac{-8}{y_2-x_2}(D_2+D_4),
\nonumber\\
N_{A_2}&=& \frac{-8}{x_2-y_2}(D_2+D_4) ,\nonumber\\
N_{A_3}&=& \frac{-8}{x_2-y_2}(D_2 - D_4).
\eea
Hard scattering amplitudes for the helicity 
$\lambda+\bar{\lambda}$(=0 or $\pm1$)
components for the diagrams $B_i(i=1,2,3)$ are given by
\bea\label{eq5}
T_{B_1}^{(\lambda+\bar{\lambda})}&=&
T_{A_2}^{(\lambda+\bar{\lambda})}( x\leftrightarrow y,
{\bf k}_\perp\leftrightarrow -{\bf l}_\perp),
\nonumber\\
T_{B_2}^{(\lambda+\bar{\lambda})}&=&
T_{A_1}^{(\lambda+\bar{\lambda})}( x\leftrightarrow y,
{\bf k}_\perp\leftrightarrow -{\bf l}_\perp),
\nonumber\\
T_{B_3}^{(\lambda+\bar{\lambda})}&=&
T_{A_3}^{(\lambda+\bar{\lambda})}( x\leftrightarrow y,
{\bf k}_\perp\leftrightarrow -{\bf l}_\perp). 
\eea

If one includes the higher twist effects such as intrinsic transverse momenta
and the quark masses, the LF gauge part proportional to
$ 1/ k^+_g$ leads to  a singularity although the Feynman gauge part
$g_{\mu\nu}$ gives the regular amplitude. This is due to the gauge-invariant
structure of the amplitudes.
The covariant derivative $D_\mu = \partial_\mu + igA_\mu$ makes both the
intrinsic transverse momenta, ${\bf k}_\perp$ and
${\bf l}_\perp$, and the transverse gauge degree of freedom $g{\bf A}_\perp$
be of the same order, indicating the need of the higher Fock state contributions
to ensure the gauge invariance.
However, we can show that 
the sum of six diagrams for the LF gauge part($1/k^+_g$ terms)
vanishes in the limit that the LF energy differences 
$\Delta_x$ and $\Delta_y$ go to zero, where $\Delta_x$ and 
$\Delta_y$ are given by
\bea\label{zb}
\Delta_x &=& M^2 - \frac{{\bf k}^2_\perp + m^2_1}{x_1}
- \frac{{\bf k}^2_\perp + m^2_2}{x_2}
= M^2 - M^2_{0x} \nonumber\\
\Delta_y &=& M^2 - \frac{{\bf l}^2_\perp + m^2_1}{y_1}
- \frac{{\bf l}^2_\perp + m^2_2}{y_2}
= M^2 - M^2_{0y}.
\eea
In the Appendix, we briefly summarize the proof.

In this work, we calculate the higher twist effects
in the limit of $\Delta_x = \Delta_y = 0$ to avoid 
the involvement of the higher Fock state contributions.
Our limit
$\Delta_x=\Delta_y=0$(but $\sqrt{\la{\bf k}^2_\perp\ra}=\beta\neq 0$)  
may be considered as a zeroth order
approximation in the expansion of a scattering amplitude. That is, 
the scattering amplitude $T_H$ may
be expanded in terms of LF energy difference $\Delta$ as 
$T_H = [T_H]^{(0)}
+ \Delta[T_H]^{(1)} + \Delta^2[T_H]^{(2)}+\cdots$, 
where $[T_H]^{(0)}$ corresponds to 
the amplitude in the zeroth order of $\Delta$. 
This approximation should be distinguished from the
zero-binding(or peaking) approximation that corresponds
to $M=m_1 + m_2$ and ${\bf k}_\perp=\beta=0$.
The point of this distinction is to note that $[T_H]^{(0)}$ includes the 
binding energy effect(i.e. ${\bf k}_\perp,{\bf l}_\perp\neq 0$) 
that was neglected in the peaking approximation.

In zeroth order of $\Delta_x$ and $\Delta_y$, the net contribution from
the LF gauge part($1/k^+_g$ terms) vanishes[see Appendix] and
we only need to compute the Feynman gauge($g^{\mu\nu}$) part, i.e.
$N_A$ and $N_B$, for the PQCD analysis of meson form factor.
The contribution of the Feynman gauge to the diagrams $A_i$ is given by
\begin{widetext}
\bea\label{THH0}
T^{(\lambda+{\bar\lambda})}_{A}&=&
\sum_{i=1}^{3}T^{(\lambda+{\bar\lambda})}_{A_i}=
N^{(\lambda+{\bar\lambda})}_A
\biggl\{
\frac{\theta(y_2-x_2)}{y_2-x_2}\frac{1}{D_1 D_2}
+ \frac{\theta(x_2-y_2)}{x_2-y_2}
\biggl(\frac{1}{D_3 D_4}+ \frac{1}{D_5 D_6}\biggr)
\biggr\}
\nonumber\\
&=&\frac{N^{(\lambda+{\bar\lambda})}_A}{(y_2-x_2)D_1D_2}\biggl\{
\theta(y_2-x_2) + \theta(x_2-y_2)(1+\frac{\Delta_{xy}}{D_{12}})
(1-\frac{\Delta_{xy}}{D_2})^{-1}(1+\frac{\Delta_y}{D_{12}})^{-1}
\biggr\},
\eea
\end{widetext}
where $D_{12} = D_1 - D_2$ and $\Delta_{xy}= \Delta_x + \Delta_y$. In the
zeroth order of $\Delta_x$ and $\Delta_y$(i.e. $\Delta_x=\Delta_y=0$), 
Eq.~(\ref{THH0}) reduces to
\bea\label{THH0_A}
T^{(\lambda+{\bar\lambda})}_{A}&=&
\frac{N^{(\lambda+{\bar\lambda})}_A}{(y_2-x_2)D_1D_2}.
\eea
Similarly for the diagrams $B_i$, we obtain 
\bea\label{THH0_B}
T^{(\lambda+{\bar\lambda})}_{B}&=&
T^{(\lambda+{\bar\lambda})}_A
(x\leftrightarrow y, {\bf k}_\perp\leftrightarrow -{\bf l}_\perp).
\eea

The hard scattering amplitude for each helicity is summarized as follows:
\begin{widetext}
\bea\label{THH}
T^{(0)}_H &=& \frac{N^{(0)}_A}{(y_2 - x_2)D_1D_2} +
(x\leftrightarrow y, {\bf k}_\perp\leftrightarrow -{\bf l}_\perp)
\nonumber\\
&=&\frac{8}{y_1}\frac{x_2y_1y_2(x_2{\bf q}^2_\perp 
+ 2{\bf k}_\perp\cdot{\bf q}_\perp) 
+ x_2(x_1 y_1 + x_2 y_2){\bf l}_\perp\cdot{\bf q}_\perp 
+ x_1y_1 m^2_2 + x_2 y_2 m^2_1 + m_1m_2(y_1-x_1)(y_2-x_2)}
{(x_2{\bf q}^2_\perp + 2{\bf k}_\perp\cdot{\bf q}_\perp)
[x_2 y^2_2(x_2{\bf q}^2_\perp + 2{\bf k}_\perp\cdot{\bf q}_\perp)
- 2x^2_2 y_2{\bf l}_\perp\cdot{\bf q}_\perp + (y_2-x_2)^2 m^2_2]}
\nonumber\\
&&+ (x\leftrightarrow y, {\bf k}_\perp\leftrightarrow -{\bf l}_\perp),
\nonumber\\
T^{(\pm)}_H &=& \frac{N^{(\pm)}_A}{(y_2 - x_2)D_1D_2} +
 (x\leftrightarrow y, {\bf k}_\perp\leftrightarrow -{\bf l}_\perp)
\nonumber\\
&=&\frac{8}{y_1}\frac{
 x_2{\bf l}_\perp\cdot{\bf q}_\perp + x_2 y_2 m^2_1 + x_1 y_1m^2_2
+ m_1m_2(y_1-x_1)(y_2-x_2)}
{(x_2{\bf q}^2_\perp + 2{\bf k}_\perp\cdot{\bf q}_\perp)
[x_2 y^2_2(x_2{\bf q}^2_\perp + 2{\bf k}_\perp\cdot{\bf q}_\perp)
- 2x^2_2 y_2{\bf l}_\perp\cdot{\bf q}_\perp + (y_2-x_2)^2 m^2_2]}
+ (x\leftrightarrow y, {\bf k}_\perp\leftrightarrow -{\bf l}_\perp),
\eea
\end{widetext}
where we neglect the terms such as ${\bf k}^2_\perp/{\bf q}^2_\perp$,
${\bf l}^2_\perp/{\bf q}^2_\perp$, and
${\bf k}_\perp\cdot{\bf l}_\perp/{\bf q}^2_\perp$ 
both in the energy denominators and the numerators 
due to the fact that ${\bf k}^2_\perp\ll{\bf q}^2_\perp$ and 
${\bf l}^2_\perp\ll{\bf q}^2_\perp$ in large momentum transfer region where
PQCD is applicable~\cite{HW}.
In the hard scattering amplitudes given by Eq.(\ref{THH}), 
the time-ordered $\theta$ 
function disappears via $\theta(x-y)+\theta(y-x)=1$ and 
there is no singularity in timelike region. 
We also note that the helicity flip contributions, i.e. 
$m_1m_2(y_1-x_1)(y_2-x_2)$, in the numerators and the mass terms 
$(y_2-x_2)^2 m^2_2$ in the denominators in Eq.~(\ref{THH}) give negligible
contributions.
One can easily find that our result  
$T^{(0)}_H$ in the leading twist limit reproduces  
the usual leading twist PQCD result, 
i.e. $T^{(0)}_H=16/(x_2y_2Q^2)$.

\section{Timelike form factor of a heavy pseuscalar meson}
We now consider the PQCD analysis of the timelike form factor
for the process of $e^+e^-$ annihilations into two pesudoscalar mesons.
The hard contribution to the timelike form factor for the electron-positron 
annihilations into two pseudoscalar mesons, i.e. $e^+e^-\to M\bar{M}$, 
is obtained as
\bea\label{FLFQM}
F_M(q^2)= e_1 I(q^2,m_1,m_2)+ e_2 I(q^2,m_2, m_1),
\eea
with the amplitude $I(q^2,m_1,m_2)$ given by 
\bea\label{TF}
I(q^2,m_1, m_2)&=& \pi C_F\alpha_s
\int \frac{dx dy d^2{\bf k}_\perp d^2{\bf l}_\perp}{16\pi^3}
\sqrt{\frac{\partial k_z}{\partial x}}
\sqrt{\frac{\partial l_z}{\partial y}}
\nonumber\\
&&\times\phi(y,{\bf l}_\perp){\cal T}_H
\phi(x,{\bf k}_\perp),
\eea
where $\alpha_s$ is the QCD running coupling constant and $C_F(=4/3)$ is
the color factor. 
We note that all the invariant masses in ${\cal R}_0$ and 
${\cal R}_{\pm 1}$ in the
spin-orbit wave function are replaced by the physical 
meson mass to be self-consistent with the result for the hard 
scattering amplitude in zeroth order of the LF energy differences
$\Delta_x$ and $\Delta_y$, i.e. $\Delta_x, \Delta_y\to 0$.  
Again, our analysis should be clearly distinguished from the peaking
approximation (zero-binding limit or $\beta\to 0$) which leads to
the DA $\phi_M(x)$ defined by 
Eq.~(\ref{Dis}) as the $\delta$-type function with $x_i= m_i/M$. 
In our zeroth order approximation of LF energy differences, 
we consider $\beta\neq 0$
(i.e. the effect of the transverse size of the meson).
In the next section (Sec.V), we take the $\beta$ values determined 
from the analysis of the mass spectroscopy using our LFQM with the 
variational principle for the QCD-motivated effective Hamiltonian~\cite{CJ2}.

Since we neglect the ${\bf k}^2_\perp,{\bf l}^2_\perp$ and 
${\bf k}_\perp\cdot{\bf l}_\perp$ terms compared to large ${\bf q}^2_\perp$ 
as we stated below in Eq.~(\ref{THH}), we have only 
${\bf k}_\perp({\bf l}_\perp)\cdot{\bf q}_\perp$ and ${\bf q}^2_\perp$
in both numerator and denominator terms
in Eq.~(\ref{THH}). Thus, for convenience in our numerical calculation, 
we change the denominator in Eq.~(\ref{THH}) to include only even 
powers of ${\bf k}_\perp$ and ${\bf l}_\perp$. Then the numerators of the
hard scattering amplitudes $T_H^{(0)}$ and $T_H^{(\pm)}$ in Eq.~(\ref{THH})  
include both even and odd powers of ${\bf k}_\perp$ and ${\bf l}_\perp$ via the
terms $({\bf k}_\perp\cdot{\bf q}_\perp)^m$ and 
$({\bf l}_\perp\cdot{\bf q}_\perp)^m$, where $m=$integer(non-negative).
After the change, the generic form of the hard scattering amplitude may be given by
\be\label{THH2}
T_H = \frac{ N(
({\bf k}_\perp\cdot{\bf q}_\perp)^m, ({\bf l}_\perp\cdot{\bf q}_\perp)^m)}
{ D({\bf q}^{2n}_\perp,({\bf k}_\perp\cdot{\bf q}_\perp)^{2n},
({\bf l}_\perp\cdot{\bf q}_\perp)^{2n})},
\ee
where $n$ and $m$ are (non-negative) integers. In Eq.~(\ref{THH2}), 
we show only essential
terms, ${\bf k}_\perp\cdot{\bf q}_\perp$ and
${\bf l}_\perp\cdot{\bf q}_\perp$ in the numerator to explain how to obtain
the nonvanishing contributions from the ordinary and the higher helicity 
contributions.  That is, as one can see from Eqs.~(\ref{TH}) and (\ref{spin}), 
by combining $T^{(0)}_H[T^{(\pm)}_H]$ in Eq.~(\ref{THH2}) with the spin-orbit 
wave function ${{\cal R}_{0}}[{\cal R_{\pm}}]$ in Eq.~(\ref{spin})
to get ${\cal T}_H$ in Eq.~(\ref{TH}) or Eq.~(\ref{TF}), 
$T^{(0)}_H[T^{(\pm)}_H]$ should have even[odd] powers of 
$({\bf k}_\perp\cdot{\bf q}_\perp)$ and
$({\bf l}_\perp\cdot{\bf q}_\perp)$ in the numerator
since ${{\cal R}_{0}}[{\cal R_{\pm}}]$ includes even[odd] powers of 
${\bf k}_\perp$ and ${\bf l}_\perp$. As a result, we can get 
${\cal T}_H$ in Eq.~(\ref{TF}) as a function of even powers of 
${\bf k}_\perp$ and ${\bf l}_\perp$ for both ordinary and higher 
helicity contributions.
We then analytically continue to the timelike region by changing 
${\bf q}_\perp$ to $i{\bf q}_\perp$(or ${\bf q}^2_\perp \to -q^2$ in this case)
in the form factor.

In terms of $F_M(q^2)$ given by Eq.~(\ref{FLFQM}), 
the cross section of the pseudoscalar meson 
pair($M\bar{M}$) production in the unpolarized $e^+e^-$ annihilations is
given by
\be\label{cross}
\frac{d\sigma}{d\Omega}(e^+e^-\to M\bar{M})=\frac{3\bar{\beta}^3}{32\pi}
\sigma_{e^+e^-\to\mu^+\mu^-}\sin^2\theta|F_M(q^2)|^2,
\ee
where $\bar{\beta}=\sqrt{1-4M^2/q^2}$ and 
$\sigma_{e^+e^-\to\mu^+\mu^-}= \pi\alpha^2/(3E^2_{\rm beam})$ with
$E_{\rm beam}=\sqrt{q^2}/2$.

In the peaking approximation, the transverse momenta of the quark and antiquark
are neglected and the longitudinal momentum fractions are given by
$x_1 = y_1 = m_1/M$ and $x_2 = y_2 = m_2/M$ with $M=m_1+m_2$.
In this approximation, the higher helicity contribution to the hard
scattering amplitude also vanishes and the ordinary helicity contribution to
the hard scattering amplitude given by Eq.~(\ref{THH}) can be rewritten as
\bea\label{pe2}
[T^{(0)}_H]_{\rm peaking}&=&
\frac{8}{x^3_2 y_1 y^2_2}
\frac{-x^2_2 y_1 y_2 q^2 + x_1 y_1 m^2_2 + x_2 y_2 m^2_1}{q^4}
\nonumber\\
&&\;\;+ (x\leftrightarrow y)
\nonumber\\
&=& \frac{32M\gamma}{q^4}\biggl(\frac{M}{m_2}\biggr)^4
\biggl[ 1 - \frac{q^2}{4M^2}\frac{2m_2}{m_1}\biggr],
\eea
where $\gamma= m_1 m_2 / M$. 
%Our result for the hard scattering amplitude
%is consistent with the peaking approximation result,i.e.  Eq.~(11) of 
%Ref.~\cite{BJ}.

Therefore, the peaking approximation of the timelike form factor of a heavy
pseudoscalar meson is given by
\bea\label{Fpeak}
[F_M]_{\rm peaking}(q^2)&\propto&
e_1\int dx dy \delta(x_i-m_i/M)[T^{(0)}_H]_{\rm peaking}
\nonumber\\
&&\times \delta(y_i-m_i/M) + (1\leftrightarrow 2)
\nonumber\\
&\propto&
\frac{1}{q^4}\biggl\{
e_1\biggl(\frac{M}{m_2}\biggr)^4
\biggl[ 1 - \frac{q^2}{4M^2}\frac{2m_2}{m_1}\biggr]
\nonumber\\
&&+
e_2\biggl(\frac{M}{m_1}\biggr)^4
\biggl[ 1 - \frac{q^2}{4M^2}\frac{2m_1}{m_2}\biggr]
\biggr\}.
\eea
This reproduces the result obtained by Brodsky and Ji~\cite{BJ}.  
The form factor zero in this approximation occurs at
\bea\label{qbar}
\bar{q}^2&=&\frac{q^2}{4M^2}
=\frac{ \frac{m_1}{2m_2} + \biggl(\frac{e_2}{e_1}\biggr)
\frac{m^3_2}{2m^3_1}}{1 + \biggl(\frac{e_2}{e_1}\biggr)\frac{m^2_2}{m^2_1}}.
\eea
Even though $m_1 > m_2$, the $e_2$ contribution is not
negligible for the heavy-heavy pseudoscalar meson system such as
$B_c$. The reason for this is because the timelike form factor of 
heavy pseudoscalar meson encounters a zero and $e_2$ contribution 
in the region near the form factor zero has non-negligible effect. 
However, for the heavy-light quark system such as $B$ and $D$ mesons, 
the light quark contribution(i.e. $e_2$) can be safely neglected.

\section{Numerical Results}
\begin{table*}[t]
\caption{The constituent quark masses $m_{q}$[GeV] and the gaussian
paramters $\beta(=\sqrt{\la{\bf k}^2_\perp\ra})$[GeV] 
for the linear potential obtained from the variational principle.
$q$=$u$ and $d$.}\label{t71}
\begin{tabular}{|c|c|c|c|c|c|c|c|c|c|c|c|c|}\hline
$m_{q}$ & $m_{s}$ & $m_{c}$ & $m_{b}$ & 
$\beta_{q\bar{q}}$ & $\beta_{s\bar{s}}$ & $\beta_{q\bar{s}}$
& $\beta_{q\bar{c}}$ & $\beta_{s\bar{c}}$ & $\beta_{c\bar{c}}$
& $\beta_{q\bar{b}}$ & $\beta_{s\bar{b}}$ & $\beta_{b\bar{b}}$\\
\hline
0.22 & 0.45 & 1.8 & 5.2 & 0.3659  & 0.4128  & 0.3886  &  0.4679 
& 0.5016 & 0.6059 & 0.5266 & 0.5712 & 1.1452\\ 
\hline
\end{tabular}
\end{table*}

In our numerical calculations, we use the model parameters
$(m_{Q{\bar Q}},\beta_{Q{\bar Q}})$ obtained from the
meson spectroscopy with the variational principle in our LFQM~\cite{CJ2}
for the linear confining potential.
Our model parameters are summarized in Table I. 
As mentioned earlier,
we should note that the r.m.s value of the transverse momentum in our LFQM
is equal to the gaussian $\beta$ value, i.e.
$\sqrt{\la{\bf k}^2_\perp\ra_{Q\bar{Q}}}=\beta_{Q\bar{Q}}$.

\begin{figure}
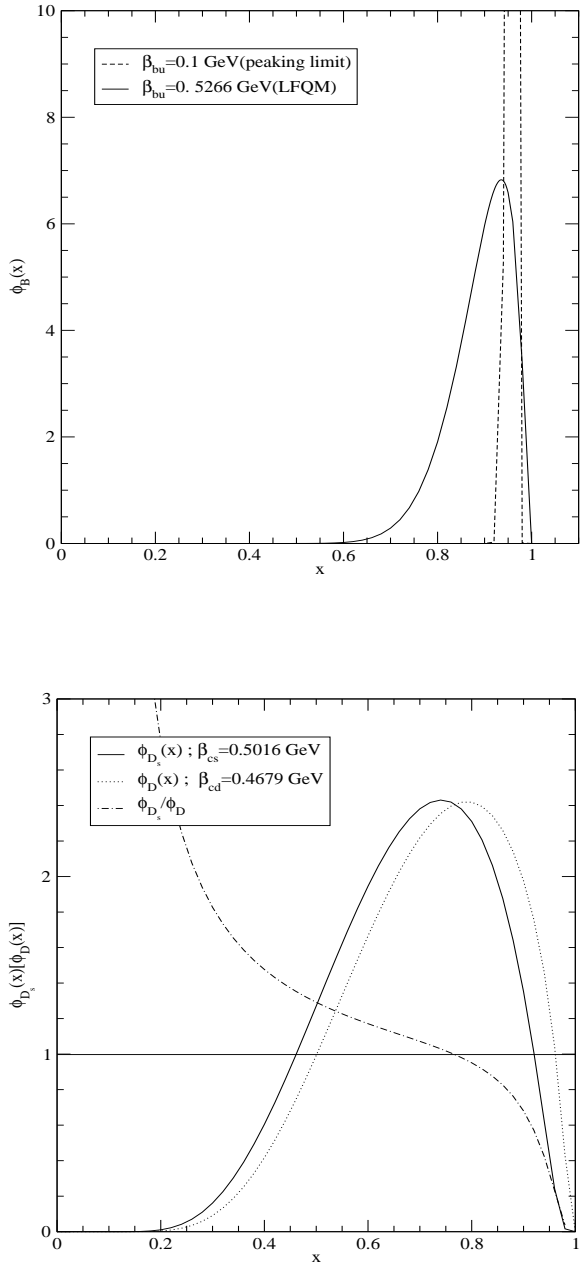

\vspace{0.8cm}
\centerline{\includegraphics[height=3in,width=3in]{Dis_B.eps}}
\vspace{1.5cm}
\centerline{\includegraphics[height=3in, width=3in]{Dis_D.eps}}
\caption{Normalized quark distribution amplitudes of $B$[top] and 
$D_s[D]$[bottom] mesons for different values of the gaussian 
parameter $\beta$.}
\label{fig2}
\vspace{0.8cm}
\end{figure}

The shape of the quark DA which depends on the $\beta$ value is
important to the calculation of the cross section for the heavy meson
pair production in $e^+e^-$ annihilations. 
We show in Fig.~\ref{fig2} the normalized quark DA of 
$B$ and $D_s[D]$ mesons with different values of $\beta$.  
For the quark DA of $B$-meson, we compare our LFQM result(solid line) with
the small $\beta$ value result close to the peaking approximation, e.g.  
$\beta=0.1$ GeV(dashed line).  As one can see, our LFQM result for the 
quark DA of $B$ meson shows sizable deviations from the peaking 
approximation.  For the quark DA of $D_s$[solid line] and 
$D$[dotted line] mesons, the peak for $\phi_D(x)$ is located to the right
of that for $\phi_{D_s}(x)$. This indicates that the $c$-quark carries 
more longitudinal momentum fraction in $D$ than in $D_s$ as one may expect. 
We also show the ratio[dash-dotted line] of $\phi_{D_s}(x)$ and $\phi_D(x)$
for the sake of comparison.

\begin{figure}
\vspace{0.8cm}
\includegraphics[height=3in,width=3in]{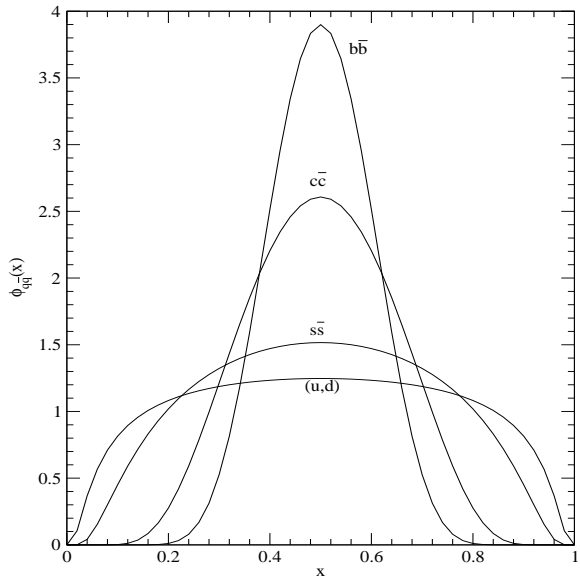}
\caption{Normalized quark distribution amplitudes of $q\bar{q}$ 
with our LFQM parameters $\beta_{q\bar{q}}$.}
\label{fig3}
\vspace{0.5cm}
\end{figure}
In Fig.~\ref{fig3}, we present also the normalized DA 
of various $^1S_0$ quarkonium($q\bar{q}$) states obtained from our LFQM 
parameters given in Table I.  
%Although we do not analyze in this work
%the cross section 
%for the $e^+e^-$ annihilation into double charmoinum production, i.e. 
%$e^+e^-\to c\bar{c} + c\bar{c}$, we intend to compare our LFQM results 
%for quark DA with those obtained from 
To explain the discrepancy between the NRQCD
prediction~\cite{Bra1} and the Belle measurement~\cite{Abe} for 
$\sigma(e^+e^-\to J/\psi + \eta_c)$, Bondar and Chernyak(BC)~\cite{BC}
reasoned that the discrepancy 
may be due to the extreme $\delta$-function-like charmonium DA 
adopted from NRQCD and claimed that they can fit the Belle data
by choosing a rather broad DA for the charmonium state. 
Interestingly, our LFQM prediction for $\phi_{c\bar{c}}$ shown in 
Fig.~\ref{fig3} looks quite similar to BC's result in a sense that the DAs
for heavy quarkonium states differ from the $\delta$-function-like DA.
In our model calculation, the DA gets narrower as
$\beta$ gets smaller. Also the timelike form factor $F_{M}(q^2)$
with small $\beta$ value decreases faster than that with large $\beta$ 
value. Since the cross section $\sigma_{e^+e^-\to M\bar{M}}$ is proportional 
to $|F_M(q^2)|^2$, the cross section with small $\beta$ is small compare 
to that with large $\beta$.  Thus, the cross section for 
$e^+e^-\to M\bar{M}$ can be in principle enhanced by broadening the quark DA.

\begin{figure}[t]
\vspace{0.8cm}
\centerline{\includegraphics[height=3in,width=3in]{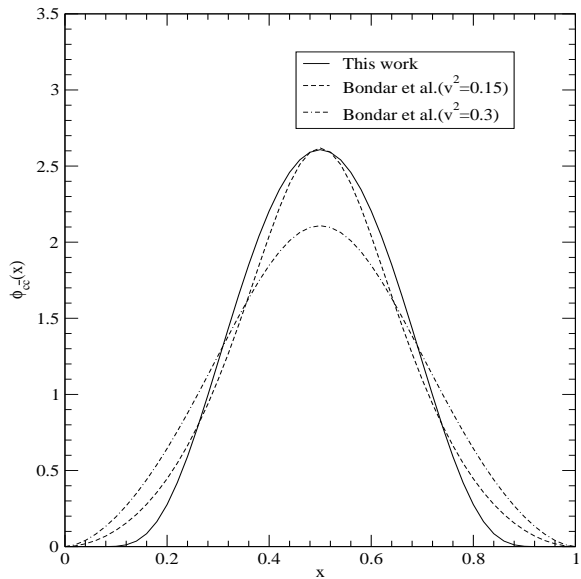}}
\caption{Normalized quark distribution amplitude of $c\bar{c}$(solid line) 
compare to those(dashed, dash-dotted lines) obtained from Bondar and 
Chernyak~\cite{BC}.}
\label{fig4}
\vspace{0.8cm}
\end{figure}

In Fig.~\ref{fig4}, we compare the results of $\phi_{c\bar{c}}(x)$
in more detail. The solid and dashed(dash-dotted) lines represent 
our LFQM result and BC's result~\cite{BC} with $v^2=0.15(v^2=0.3)$, 
respectively, where the parameter $v$\cite{BC} represents the characteristic quark velocity 
in the bound state. Although there is a similarity in the quark DA between 
ours and BC's results(in particular, their $v^2=0.15$ result), there is a rather 
substantial difference near the end-point region between ours and BC's results. 
Since the PQCD hard scattering amplitude is typically very sensitive 
to the end-point values of DA, it may not be so difficult to imagine
that BC's prediction of double charm production cross sections
would have made a difference depending on what $v^2$ value they have used.
To fit the Belle measurement~\cite{Abe} for $\sigma(e^+e^-\to J/\psi + \eta_c)$,
they used $v^2=0.3$ rather than $v^2=0.15$.
However, as we stressed earlier, the way that BC\cite{BC} handled the 
transverse momentum effect is different from ours
since they integrated out the transverse momentum seperately in the 
wave function part and in the hard scattering part while we didn't 
factorize the hard and soft parts but integrated out the transverse momentum
for the whole amplitude. 
We think that a consistent analysis with $\beta\neq 0$ 
(or non-zero binding energy) should follow our non-factorized 
formulation (see e.g. Eq.(\ref{TF})).

\begin{figure}[t]
\vspace{0.9cm}
\centerline{\includegraphics[height=3in,width=3in]{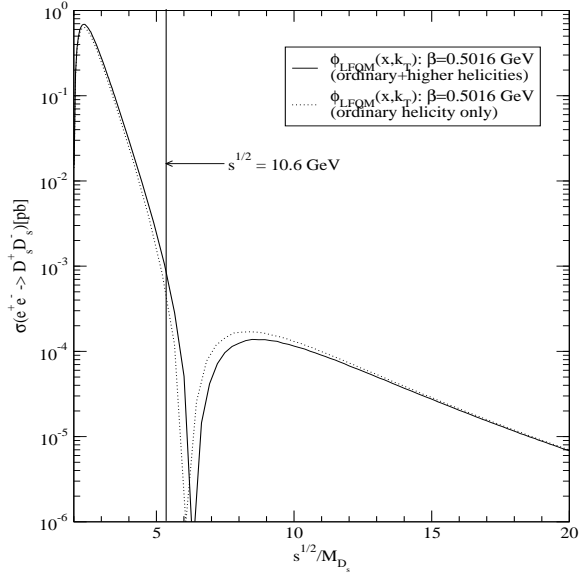}}
\vspace{1.5cm}
\centerline{\includegraphics[height=3in,width=3in]{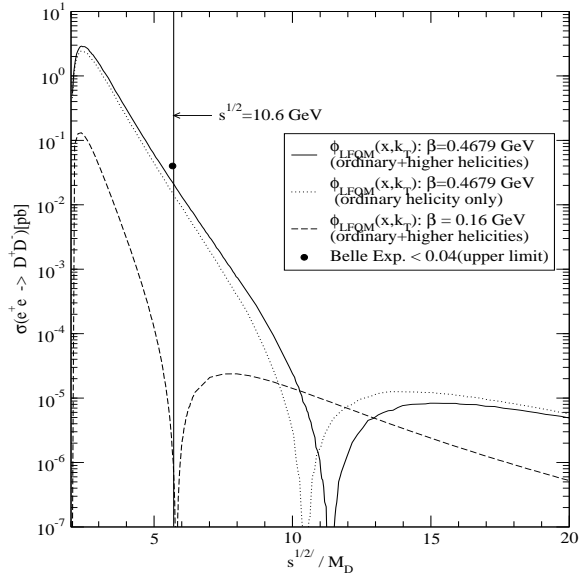}}
\caption{Cross sections for $D_s^+D_s^-$[top] and $D^+D^-$[bottom]
 pair productions in $e^+e^-$ annihilations with $\alpha_s=0.2$.}
\label{fig5}
\vspace{0.8cm}
\end{figure}

In Fig.~\ref{fig5}, we show the cross sections for the exclusive $D_s^+D_s^-$ 
and $D^+D^-$ pair productions in $e^+e^-$ annihilations.
The solid line represents the results including both ordinary and 
higher helicity contributions while the dotted line corresponds to 
the result of the ordinary helicity contribution only. 
The dashed line for $\sigma(e^+e^-\to D^+D^-)$ represents the lower
limit(i.e. $\beta=0.16$ GeV) for the form factor zero to occur above
$\sqrt{s}=10.6$ GeV. 
The small black circle for $\sigma(e^+e^-\to D^+D^-)$ represents the
upper limit obtained from Belle~\cite{Uglov},
i.e. $\sigma_{\rm Exp.}(e^+e^-\to D^+D^-)<0.04$.
The higher helicity contribution 
to the cross section, i.e. the difference between solid and dotted line, 
is more pronounced in $e^+e^-\to D^+D^-$ than in $e^+e^-\to D_s^+D_s^-$ 
especially near the turnover point (or the form factor zero point). 
In general, the higher helicity contribution to the meson 
form factor increases(decreases) as quark mass decreases(increases).
For instance, while the higher helicity contribution to the hard scattering
amplitude is negligible in the $B$ meson form 
factor, it is not negligible in the pion form factor. 
However, the most significant in our analysis is
the transverse momentum effect which delays the
turnover point (see Eq.~(\ref{qbar}) for the peaking approximation).
For instance, the turnover for $D_s[D]$ meson occurs near 6.3[11.3] 
by going beyond the peaking approximation while the corresponding 
turnover point is near $\sqrt{s}/M\sim 2.8[4.03]$ for the 
peaking approximation.

Numerically, we obtain the cross sections for $D_s^+D_s^-$ and 
$D^+D^-$ pair productions at $\sqrt{s}=10.6$ GeV with our LFQM
paramters(i.e. $\beta=0.5016$ GeV for $D_s$ and 0.4679 GeV for $D$) 
as follows
\bea
\sigma(e^+e^-\to D_s^+D_s^-)&=& 
(8.0^{+4.4}_{-3.5})\times 10^{-4}[\rm pb],
\nonumber\\
\sigma(e^+e^-\to D^+D^-)&=& (0.02\pm 0.01)[\rm pb],
\eea
for the strong coupling constant $\alpha_s=0.2\pm 0.05$. 
Similar $\alpha_s$ values were used in Refs.~\cite{BSJ,KL}.
Our result for $\sigma(e^+e^-\to D^+D^-)$
is consistent with the recent experimental data from Belle,
$\sigma_{\rm Exp.}(e^+e^-\to D^+D^-)<0.04$.
Since $\sigma(e^+e^-\to D^+D^-)$ gets larger as $\beta$ grows, 
the upper bound of $\sigma_{\rm Exp.}$ from Belle provides a constraint
on the maximum $\beta$ value modulo the dependence on $\alpha_s$.

\begin{figure}
%\vspace{0.9cm}
\centerline{\includegraphics[height=3in,width=3in]{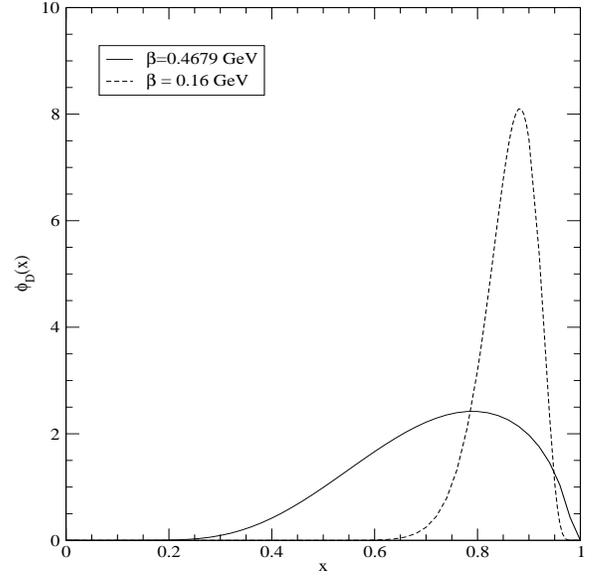}}
\caption{Lower bound for the shape of $\phi_D(x)$[dashed line] deduced from 
the assumption of $\sigma(e^+e^-\to D^+D^-)<0.04$[pb] 
and $(d\sigma/d\sqrt{s}) <0$ at $\sqrt{s}=10.6$.}
\label{fig6}
\vspace{0.8cm}
\end{figure}

If the cross section for the $D$ meson pair production
satisfies $\sigma(e^+e^-\to D^+D^-)<0.04$ 
and its slope with respect to the momentum transfer is negative, i.e.
$(d\sigma/d\sqrt{s}) <0$ at $\sqrt{s}=10.6$ GeV, then we could also set the
lower bound for $\beta$ value as $\beta\geq 0.16$.
The shape of the $D$ meson quark DA corresponding to $\beta = 0.16$ GeV
is shown by dashed line in Fig.~\ref{fig6}. 
If $\beta<0.16$ GeV, the $D$ meson quark DA approaches to the $\delta$-type 
function but $(d\sigma/d\sqrt{s}) >0$ at $\sqrt{s}=10.6$ GeV due to the form 
factor zero occuring at $\sqrt{s}< 10.6$ GeV. Due to the occurence of form 
factor zero for the heavy pseudoscalar meson pair production~\cite{Grozin,BJ,BCS}, 
more experimental data around $\sqrt{s}=10.6$ GeV are necessary to check 
the slope of the cross section. More data around $\sqrt{s}=10.6$ GeV 
would further constrain the shape of $D$ meson quark DA. 

\begin{figure}
%\vspace{0.9cm}
\centerline{\includegraphics[height=3in,width=3in]{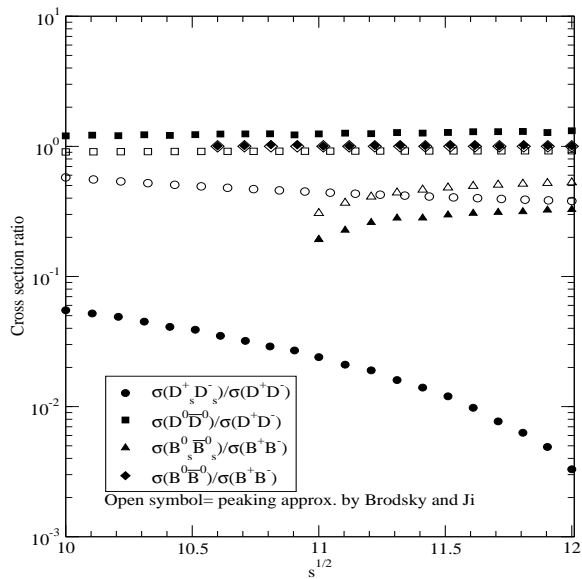}}
\caption{Our predictions[closed symbols] on the cross section ratios
for various heavy pseudoscalar meson($B,B_s,D$,and $D_s$) pair productions
compared to the peaking approximation
results[open symbols] near $\sqrt{s}=10.6$ GeV.}
\label{fig7}
\vspace{0.8cm}
\end{figure}

How about $B$ mesons? We should point out that the PQCD
result of the cross section for $B^+ B^-$ pair production may not be
trustworthy because $\sqrt{s} = 10.6$ GeV is too close to the threshold
energy of $B^+ B^-$ pair production. As expected,
the gluon momentum transfer from the heavy quark to the light quark in
$B^+ B^-$ pair production at $\sqrt{s} = 10.6$ GeV turns out to be only
around a few hundred MeV close to the scale of $\Lambda_{QCD}$.
On the other hand, the gluon momentum transfer in $D_{[s]}^+ D_{[s]}^-$
pair production at $\sqrt{s} = 10.6$ GeV is much larger than
the scale of $\Lambda_{QCD}$.
By going beyond the peaking approximation,
the average gluon momentum transfer gets even larger due to the transverse
momentum effect. 
This may justify our PQCD analysis for $D_{[s]}^+ D_{[s]}^-$
pair production at $\sqrt{s} = 10.6$ GeV. 

Although the absolute value of
the cross section for $B[B_s]$-meson may not be reliable 
near $\sqrt{s} = 10.6$ GeV, it seems interesting to discuss the behavior
of the ratios of cross sections such as 
$\sigma(e^+e^-\to B^0\bar{B}^0)/\sigma(e^+e^-\to B^+B^-)$ and
$\sigma(e^+e^-\to B_s^0\bar{B_s}^0)/\sigma(e^+e^-\to B^+B^-)$.
In Fig.~\ref{fig7}, we show our predictions[closed symbols] on 
the cross section ratios 
for various heavy pseudoscalar meson($B,B_s,D$,and $D_s$) pair productions
in $e^+e^-$ annihilations near $\sqrt{s}=10.6$ GeV, i.e.$\sqrt{s}$ ranging 
from 10 to 12 GeV and compare our results with those[open symbols] 
obtained from the peaking approximation~\cite{BJ}.
The open and closed diamond symbols for
$\sigma(e^+e^-\to B^0\bar{B}^0)/\sigma(e^+e^-\to B^+B^-)$ are on top of each 
other and their values are almost equal to 1. 
In fact, the cross 
section ratios involving light quarks $u$ and $d$ such as $(D^0, D)$ and 
$(B^0, B)$ cases are close to 1.  This is due to the negligible contribution
from the diagrams where the photon is attached to the light quarks.
As the replacement of light quarks by the strange quark makes those diagrams
non-negligible, the cross section ratios 
for the cases of $(D_s, D)$ and $(B_s, B)$ deviate from 1 appreciably.
However, most significant is again the transverse momentum effect which is
pronounced in the case of $D[D_s]$ meson pair productions 
compared to $B[B_s]$ meson pair productions. In particular, the deviation between
the open and closed symbols for the case $(D_s, D)$ is quite dramatic
compared 
to the case of $(B_s, B)$.
\section{Summary and Conclusion}
We investigated the transverse momentum effect on the exclusive heavy 
meson pair productions in $e^+e^-$ annihilations within the framework of 
LF PQCD. The gaussian parameter $\beta$ in our model wave function is 
found to be related to the transverse momentum via
$\beta_{Q\bar{Q}}=\sqrt{\la{\bf k}^2_\perp\ra_{Q\bar{Q}}}$. This relation
naturally explains the zero-binding energy limit for the zero transverse
momentum, i.e.  $\la M^2_0\ra= (m_1 + m_2)^2$ and $x_i=m_i/M$
for $\beta= 0$. 

However, the heavy quark DA is sensitive to the value of $\beta$
and indeed substantially broad and quite different from the $\delta$-type DA  
according to our LFQM based on the variational principle for 
the QCD-motivated Hamiltonian~\cite{CJ2,CJ99}. 
If the quark DA is not an exact $\delta$ function, i.e. 
${\bf k}_\perp$ in the soft bound state LF wave function 
can play a significant role, the factorization theorem is no longer applicable. 
To go beyond the peaking approximation, the invariant amplitude
should be expressed in terms of the LF wave function 
$\Psi(x_i,{\bf k}_{\perp i})$ rather than the quark DA.

In going beyond the peaking approximation, we stressed a consistency 
by keeping the transverse momentum ${\bf k}_\perp$ both in the wavefunction part 
and the hard scattering part together before doing any integration 
in the amplitude. Such non-factorized analysis should be distinguished from the 
factorized analysis where the transverse momenta are seperately integrated out in the
wavefunction part and in the hard scattering part.
Even if the used LF wavefunctions lead to the similar shapes of DAs, 
predictions for the cross sections of heavy meson productions would 
apparently be different between the factorized and non-factorized analyses.

In this work, we compared our non-factorized analysis
with the usual factorized analysis based on the peaking approximation
and found a substantial difference between the two in the calculation of 
the cross section for the heavy meson pair production.
We also discussed the higher helicity contribution to the cross section.  
Our analysis provided a constraint on the size of quark transverse momentum
inside the $D$ meson from the recent Belle data,
$\sigma_{\rm Exp.}(e^+e^-\to D^+D^-)<0.04$. 
More experimental data around $\sqrt{s}=10.6$ GeV would further constrain 
the shape of $D$ meson quark DA and test our LFQM prediction. 
Application of our non-factorized PQCD analysis to the $alpha_s$ higher order 
corrections, e.g. in double charm production, would deserve further investigation.

\acknowledgements
This work was supported by a grant from the U.S. Department of 
Energy(DE-FG02-96ER40947). H.-M. Choi was supported in part by Korea Research 
Foundation under the contract KRF-2005-070-C00039.
The National Energy Research Scientific Center is also acknowledged for the
grant of supercomputing time. 

\appendix
\section{ Proof of vanishing light-front gauge part in 
$M=M_0$ limit}
\label{AppA}
The contribution of the LF gauge part($1/k^+_g$ term) to the hard scattering
amplitude for diagrams $A=\sum_{i=1}^{3}A_i$ obtained from 
Eqs.~(\ref{eq1}) and (\ref{eq3}) is given by
\begin{widetext}
\bea\label{LG}
T^{(1/k^+_g)}_{A}&=& 
\frac{\theta(y_2-x_2)}{y_2-x_2}\frac{N_{A_1}}{D_1 D_2}
+ \frac{\theta(x_2-y_2)}{x_2-y_2}
\biggl(\frac{N_{A_2}}{D_3 D_4}+ \frac{N_{A_3}}{D_5 D_6}\biggr)
\nonumber\\
&=&-\frac{8}{(y_2-x_2)^2}\frac{D_2+D_4}{D_1D_2}\biggl\{
\theta(y_2-x_2) + \theta(x_2-y_2)\biggl[
1+ \frac{D_2-D_4}{D_4}\frac{D_1D_2 + D_5(D_2+D_4)}{(D_2+D_4)D_5}
\biggr]\biggr\}.
\eea
\end{widetext}
In terms of the LF energy differences,
the relevant energy denominators can be rewritten as
\bea\label{Di}
D_1 &=& \Delta_x
- \frac{1}{x_1}(x_2{\bf q}^2_\perp + 2 {\bf k}_\perp\cdot{\bf q}_\perp)
\nonumber\\
D_2+D_4 &=& D_1 + \Delta_y
\nonumber\\
D_2+ D_5&=& \Delta_x + \Delta_y.
\eea
Eq.~(\ref{LG}) and the corresponding LF gauge part to the diagrams $B_i$
lead to singularties. Fortunately, however, in the zeroth order of 
$\Delta_x$ and $\Delta_y$, i.e. $\Delta_x, \Delta_y\to 0$ limit, one
can see that the energy denominator term in $\theta(x_2-y_2)$ in 
Eq.~(\ref{LG}) vanishes, which leads to 
$\theta(y_2-x_2)+\theta(x_2-y_2)=1$. 
Thus, the LF gauge part contribution to the diagrams 
$A=\sum_{i=1}^{3}A_i$ becomes 
\bea\label{LGA}
T^{(1/k^+_g)}_A&=& -\frac{8}{(y_2-x_2)^2}\frac{1}{D_2},
\eea 
and similarly we obtain
\bea\label{LGB}
T^{(1/k^+_g)}_B&=& -\frac{8}{(y_2-x_2)^2}\frac{1}{D_9},
\eea
for the diagrams $B=\sum_{i=1}^{3}B_i$ . 
Finally, from the relation $D_2 +D_9 = \Delta_x + \Delta_y$, one can 
see that the net contribution 
from the LF gauge parts, i.e., $T^{(1/k^+_g)}_A + T^{(1/k^+_g)}_B$, 
vanishes exactly in the limit of $\Delta_x=\Delta_y=0$.


\newpage
\begin{thebibliography}{99}
\bibitem{Aubert} BarBar Collaboration, B. Aubert {\em et al.},
\Journal{\PRL}{87}{162002}{2001}.
\bibitem{Abe} Belle Collaboration, K. Abe {\em et al.},
\Journal{\PRL}{88}{052001}{2002}; \Journal{\PRL}{89}{142001}{2002}.
\bibitem{Uglov} Belle Collaboration, T. Uglov {\em et al.},
\Journal{\PRD}{70}{071101}{2004}.
\bibitem{Grozin} A. G. Grozin and M. Neubert,
\Journal{\PRD}{55}{272}{1997}.
\bibitem{Bra1} E. Braaten and Jungil Lee, \Journal{\PRD}{67}{054007}{2003}.
\bibitem{Liu} K. Y. Liu, Z.G. He and K. T. Chao,
\Journal{\PLB}{557}{45}{2003};
K. Hagiwara, E. Kou and C. F. Qiao, \Journal{\PLB}{570}{39}{2003}.
\bibitem{Liu2} K. Y. Liu, Z. G. He, Y. J. Zhang, and K. T. Chao,
hep-ph/0311364.
\bibitem{BKL2006} G. T. Bodwin, D. Kang and J. Lee, hep-ph/0603185.
\bibitem{Cho} P. Cho and K. Leibovich, \Journal{\PRD}{54}{6990}{1996};
F. Yuan, C. F. Qiao, and K. T. Chao, \Journal{\PRD}{56}{321}{1997};
S. Baek, P. Ko, J. Lee, and H. S. Song, J. Korean Phys. Soc. {\bf 33},
97 (1998).
\bibitem{Chao2006} Y.-J. Zhang, Y.-J. Gao and K.-T.Chao, 
\Journal{\PRL}{96}{092001}{2006}.
\bibitem{BJ} S. J. Brodsky and C.-R. Ji,
\Journal{\PRL}{55}{2257}{1985}.
\bibitem{BL} G. P. Lepage and S. J. Brodsky, \Journal{\PRD}{22}{2157}{1980}.
\bibitem{JP} C.-R. Ji and A. Pang, \Journal{\PRD}{55}{1253}{1997}.
\bibitem{BC} A. E. Bondar and V. L. Chernyak, \Journal{\PLB}{612}{215}{2005}.
\bibitem{MS} J. P. Ma and Z. G. Si, \Journal{\PRD}{70}{074007}{2004}.
\bibitem{CJ2} H.-M. Choi and C.-R. Ji, \Journal{\PLB}{460}{461}{1999}.
\bibitem{CJ99} H.-M. Choi and C.-R. Ji, \Journal{\PRD}{59}{074015}{1999}.
\bibitem{Melosh} H. J. Melosh, \Journal{\PRD}{9}{1095}{1974}.
\bibitem{CJ_Jacob} H.-M. Choi and C.-R. Ji, \Journal{\PRD}{56}{6010}{1997}.
\bibitem{DYW} S. D. Drell and T. M. Yan, \Journal{\PRL}{24}{181}{1970};
G. West, {\em ibid.}{\bf 24}, 1206 (1970).
\bibitem{BHL} G.P. Lepage, S. J. Brodsky, T. Huang, and P. B. Mackenzie,
in {\em Particles and Fields}, Proceeding of the Banff Summer Institute on
Particle Physics, Banff, Alberta, Canada, 1982, edited by A. Z. Capri and A. N.
Kamal(Plenum, New York, 1983), p. 143. 
\bibitem{HW} T. Huang, X.-G. Wu, and X.-H. Wu, 
\Journal{\PRD}{70}{053007}{2004}.
\bibitem{BSJ} S.J. Brodsky, A.S. Goldhaber, and J. Lee,
\Journal{\PRL}{91}{112001}{2003}.
\bibitem{KL} D.Kang {\em et al.}, \Journal{\PRD}{71}{071501(R)}{2005}.
\bibitem{BCS} M.S.Baek, S.Y.Choi, and H.S.Song, \Journal{\PRD}{50}{4363}{1994}.
\end{thebibliography}
\end{document}